\documentclass[a4paper,11pt]{article}
\usepackage{pos}
\usepackage{graphicx}
\usepackage{color}

\title{X-ray Spectroscopy of Disk Winds in Black Hole X-ray Binaries}
\author*[a]{Megumi Shidatsu}
\author[a]{Maxime Parra}

\affiliation[a]{Graduate chool of Science and Engineering, Ehime University,\\
 2-5, Bunkyocho, Matsuyama, Ehime, Japan}

\emailAdd{shidatsu.megumi.wr@ehime-u.ac.jp}
\emailAdd{maxime.parrastro@gmail.com}

\abstract{Powerful outflows along the accretion disk, known as disk winds, are sometimes launched in black hole X-ray binaries. These winds often manifest themselves in X-ray spectra as blueshifted, highly ionized absorption lines. Previous observations suggest that the mass loss rate from the disk due to disk winds can be comparable to or even more than the mass accretion rate onto the black hole, indicating that disk winds likely play crucial roles in shaping the accretion disk structure and affecting the surrounding environment. However, the mechanisms driving these winds, as well as how their structure changes in response to variations in the mass accretion rate, remain poorly understood. The X-ray Imaging and Spectroscopy Mission (XRISM), launched in September 2023, is equipped with Resolve, a cutting-edge X-ray micro-calorimeter that delivers unprecedented spectral resolution. 
Resolve is expected to significantly advance our understanding of wind launching mechanisms and their impact on accretion processes and environments. In this article, we review the progress made in the pre-XRISM era, highlight key results obtained from XRISM observations to date, and outline future prospects.
}

\FullConference{87th Fujihara Seminar：The 50th Anniversary Workshop of the Disk Instability Model in Compact Binary Stars (DIM50TH2025)\\
22-26 September 2025\\
Tomakomai, Japan\\}

\begin{document}
\maketitle

\section{Introduction}
\label{sec:intro}

Outflows are present in accreting compact objects over a wide range of mass scales, including supermassive black holes and stellar mass compact objects. Observationally, they are categorized into two different types: highly collimated jets launched almost perpendicular to the accretion disk, and uncollimated disk winds mainly extending along the disk plane. These outflow can be so powerful and affect the disk structure and thereby the disk instability and accretion processes. 
They may also have a significant impact to various areas of modern astrophysics. 
For example, powerful outflows driven by accreting supermassive black holes (or active galactic nuclei: AGN) are considered to play a key role in co-evolution of galaxies and their central black holes (e.g, \cite{fabian2012}). However, a number of fundamental questions remain unanswered: what is their main launching mechanism? How much mass and energy do they actually carry away and how does it depend on the mass accretion rate and black hole spin? 

Galactic black hole X-ray binaries (BHXBs), close binary systems  consisting of an accreting stellar-mass black hole and a donor star, are ideal targets to tackle these important questions. Many of known Galactic BHXBs,  
especially hosting low-mass companion stars, show a transient nature. They are usually in a quiescent phase but suddenly cause an outburst and increase there X-ray luminosity by several orders of magnitude in a week or so, and then gradually dim for several weeks to a few year \cite{mcclintock2006,done2007,tetarenko2017}. 
Because the X-ray luminosity approximately traces the mass accretion rate, monitoring outbursts enables us to investigate the evolution of the accretion flows and outflows over a wide range of the mass accretion rates. Compared with AGN, Galactic BHXBs can reach much higher apparent brightness during their outbursts because of their proximity. Moreover, their variation occurs on much shorter time scales, making them well suited for detailed observational studies and allowing us to track the evolution of outflows across a broad range of mass accretion rates.

Disk winds in black hole X-ray binaries (BHXBs), the main topic of this article, have been observed as blueshifted, ionized absorption lines in the X-ray band and some other wavelengths. In X-rays, they usually exhibit lines from highly ionized ions, especially H and He-like irons (e.g., \cite{ueda1998,lee2002,miller2006,kubota2007,diaztrigo2014,hori2018,parra2025}). 
%They are only seen in high-inclination systems, suggesting that the winds have an equatorial structure, extending along the disk plane with a small solid angle (e.g., \cite{ponti2012}). 
As described below, they often carry away so much mass that they can affect the disk structure and may have a significant impact on the conditions of disk instability and subsequent accretion processes.  
In addition, the observed spectral signature of winds depend on the spectral states \cite{ponti2012,parra2024}, which are considered to trace the structure of the accretion disk in the vicinity of the black hole, as well as the presence and absence of jets. Studying  winds are hence important to figure out the whole picture of the black hole accretion and outflows and their effects to the black hole and environment.

The X-ray Imaging and Spectroscopy Mission (XRISM; \cite{tashiro2025}), launched in September 2023, carries the state-of-the-art X-ray microcalorimeter Resolve \citep{kelley2025,ishisaki2025}, which provides unprecedented spectral resolution and is expected to significantly advance our understanding of winds in BHXBs. In this proceeding, we review X-ray studies of winds conducted prior to the launch of XRISM, present the latest results from XRISM observations, and discuss future prospects.

\section{Review of X-ray Observations and Interpretations of Winds in BHXBs: before XRISM Era}
\label{sec:review}
%% history of x-ray observations of winds 
%%% detections from ASCA era, physical parameters determined by observables

\subsection{X-ray Spectroscopy of Winds}

Ionized Fe K absorption lines from Galactic BHXBs were first detected in ASCA observations of GRO J1655$-$40 \citep{ueda1998, yamaoka2001} and GRS 1915$+$105 \citep{kotani2000}. Subsequently, similar features were detected in neutron star low-mass X-ray binaries, such as GX 13$+$1 (e.g., \citep{ueda2001,sidoli2002}), suggesting that ionized absorption is ubiquitously present, 
not depending on the existence or absence of a solid surface. Since then, absorption features have been detected in $\sim 10$ BHXBs \citep{parra2024}, often exhibiting significant blueshifts that provide clear evidence for outflowing winds. These systems have a relatively high inclination angles ($\gtrsim 60^\circ$), and therefore the winds are thought to have an equatorial geometry, extending along the disk plane with a limited solid angle \citep{ponti2012,parra2024}. In addition, the absorption lines are persistently observed throughout their orbital phases (e.g. \cite{yamaoka2001, sidoli2001, sidoli2002}), indicating that the wind blow out in wide directions rather than in a specific direction.

Some key physical parameters of winds can be determined from the observed absorption lines. In particular, the line-of-sight outflow velocity $v_{\rm out}$ is derived from the measured blueshifts of the lines, and has been found to be typically of the order of 100--1000 km s$^{-1}$ in previous observations (e.g., \cite{parra2024,munozdarias2026}).
Another important parameter that can be inferred from the observations is the distance of the ionized absorber $R$. This can be estimated using the  number density $n$, ionizing luminosity $L$, and ionization parameter $\xi \equiv L/nR^2$ as $R = \sqrt{L/(n\xi)}$. The density is often difficult to measure, so we modify the equation using the radial extension of the absorber $\Delta R$ and the column density $N_{\rm H} (\sim n \Delta R)$ as 
\begin{equation}
    R \sim \frac{L}{N_{\rm H} \xi} \cdot\frac{\Delta R}{R}.
\end{equation}
From typical values of $N_{\rm H} \sim 10^{22-23}$ cm$^{-1}$, $L \sim 10^{38}$ erg s$^{-1}$, and $\log\xi \sim 5$ obtained from observations, we obtain $R \sim 10^{10-11} (\Delta R/R)$ cm (e.g., \cite{kubota2007}). This indicates that winds are likely to be located in the outer region of the accretion disk, although the uncertainty of $\Delta R/R$ remains. These parameters then give an estimate of the mass outflow rate, which can be expressed as: 
\begin{equation}
  \dot{M}_{\rm out} \sim 4 \pi C \mu m_{\rm p} N_{\rm H} R v_{\rm out}, 
\end{equation}
where $m_{\rm p}$ is the proton mass, $\mu$ is the mean mass par particle, and $C$ is the covering fraction of the wind, defined with the solid angle $\Omega$ as $\Omega/4\pi$. The $\dot{M}_{\rm out}$ values were $10^{18-20}$ g s$^{-1}$, comparable to or in some cases much larger than the mass accretion rate through the inner disk region (e.g., \cite{kotani2000, miller2006, ueda2009, neilsen2011}). Therefore, understanding of the nature of disk winds is indispensable to build the whole picture of accretion disk dynamics in BHXBs. 

%% state dependence of the x-ray winds
The detectability of X-ray absorption lines shows a strong dependence on the spectral state \citep{ponti2012,parra2024}. These features are predominantly observed in the high/soft state, in which the emission is dominated by soft X-rays from the standard accretion disk. As the spectrum hardens, the absorbing material tends to become more highly ionized (e.g., \cite{diaztrigo2014,hori2018,parra2025}). In contrast, such absorption lines are generally absent in the low/hard state \citep{miller2012,neilsen2019}, where the X-ray spectrum is dominated by a hard power-law component, considered to be produced by Comptonisation of the soft X-rays in the hot flow or corona surrounding the disk. 
A possible explanation of the line disappearance is the thermal instability in the wind driven by the spectral change during the transition \citep{bianchi2017,petrucci2021}. However, how the wind properties (such as the density, ionization stage, velocity) actually change between these two states is not well understood. 
Interestingly, the behavior of wind absorption lines appears to be anti-correlated with the presence of radio jets. Steady, compact jets are commonly observed in the low/hard state, whereas they are quenched in the high/soft state \citep{fender2004}. Prior to jet quenching, powerful transient jets extending to parsec scales are sometimes launched during the state transition from the low/hard to the high/soft state, which typically occurs near the peak of an outburst. However, the physical background of the connection between the evolution of jets, the accretion disk, and winds is an open question.

\subsection{Wind Launching Mechanism}
%% wind launching mechanism
Understanding the launching mechanisms of winds is crucial for interpreting their observational properties and evaluating their impact on accretion processes and environments, yet this remains a long-standing open issue. Three possible mechanisms have been proposed so far: radiation-pressure driving, thermal driving, and magnetic driving. 

Radiation pressure driven winds are launched when the outward radiation force overcomes the gravitational force by the black hole. The radiation force can be exerted through Compton scattering (continuum driven winds) or the bound–bound/bound–free absorption (so-called line driven winds).
The continuum driven wind, by definition, is only dominant  
above the Eddington luminosity ($L_{\rm Edd}$). However, most of the known BHXBs do not reach $L_{\rm Edd}$ even at the flux peak of the outbursts.  
Pure continuum driven winds are therefore unlikely to explain the majority of the disk winds seen in BHXBs, although it could work in some exceptional cases.
Line driven winds are considered as a plausible mechanism for some winds in AGN \citep{proga2000,nomura2016}, but it is again unlikely to work in BHXBs because their much higher temperature disks mean that the strong UV absorption species are completely ionized \citep{proga2002}. 

Thermal winds are dominated by the gas pressure. In this scenario, the outer disk regions are irradiated by the strong X-rays emitted from the inner disk region, and gas in the disk photosphere is heated up to the so-called Compton temperature, $T_{\rm IC}$, where Compton up- and down-scattering are balanced. This temperature is described as follows: 
\begin{equation}
   T_{\rm IC} = \frac{1}{4k}\frac{\int_0^\infty h\nu L_\nu d \nu} {\int_0^\infty L_\nu d \nu},
\end{equation}
where $h$ is the Planck constant and $k$ is the Boltzmann constant (see e.g., \cite{begelman1983,done2010}), and hence can be determined purely by the spectral energy distribution. Its typical value for BHXBs is estimated to be $\sim 10^7$ K in the high/soft state, where the absorption lines are visible. The heated gas can escape from the disk at the radii where its kinetic energy overcomes the local gravitational energy. This gives an estimate for the wind launching radius,
\begin{equation}
 R_{\rm IC} = \frac{\mu m_{\rm p} G M_{\rm BH}}{kT_{\rm IC}},   
\end{equation} 
although it is just a rough estimate and a more careful analysis suggests  that thermal winds can be actually formed at $\gtrsim$ 0.1 $R_{\rm IC}$ \citep{begelman1983,woods1996}. The launching radius of a thermal wind is thus estimated to be $\sim 10^{10}~(T_{\rm IC}/10^7~{\rm K})^{-1}$ cm or $\sim 10^{5}~(T_{\rm IC}/10^7~{\rm K})^{-1}~(M_{\rm BH}/10~M_\odot)^{-1}~R_{\rm g}$ (where $R_{\rm g}$ is the gravitational radius, $GM_{\rm BH}/c^2$), which is consistent with the location of the absorber estimated from the observed lines. The gas expands at the sound speed, typically 100--1000 km s$^{-1}$, which is also consistent with the line-of-sight velocity of many observed winds. The irradiated gas at smaller radii than the launching radius of thermal winds remains bound on the surface of the disk, forming a hot ionized atmosphere. This can explain the ionized absorption feature without a significant blueshift seen in X-ray binaries with short orbital periods \citep{diaztrigo2016}. 

In the above calculations, the temperature of the irradiated gas was assumed to reach $T_{\rm IC}$. However, at low luminosities, Compton scattering becomes inefficient, preventing from raising the gas temperature to $T_{\rm IC}$ and thereby suppressing the launching of winds. This critical luminosity is given as $L_{\rm crit} \sim 3 \times 10^{-2}~(T_{\rm IC}/10^8~{\rm K})^{-1/2}~L_{\rm Edd}$ \citep{begelman1983}, and for typical $T_{\rm IC}$ value in the high/soft state, thermal winds are predicted to disappear below $\sim$ several $10\%~L_{\rm Edd}$. In addition, in the low/hard state, X-ray irradiation of the outer disk becomes insufficient due to self-shielding by a hot atmosphere in the inner disk region \citep{tomaru2019}. This effect, together with the harder continuum spectrum that more strongly ionizes the gas in the outer disk \citep{done2018,shidatsu2019}, could explain the absence of absorption lines in the low/hard state.
Meanwhile, at high sub-Eddington luminosities, the radiation force provides an additional but non-dominant component. Both theoretical works \citep{proga2002,done2018} and numerical simulations \citep{higginbottom2020,tomaru2019,tomaru2022} showcase a significant reduction of the launching radius and increase in outflow velocity above $\sim 0.2~L_{\rm Edd}$. 

The last type of winds is a magnetic wind, powered by the magnetic processes. Compared with thermal winds, magnetic winds are difficult to study in a quantitative manner as the magnetic field configuration around the accretion disk is not known. They are however likely to exist at some level because moderately strong magnetic fields should be present to produce sufficient viscosity for angular momentum transport \cite{balbus2002} and allow jet ejections without the requirement of high speed systems 
(e.g., \cite{blandford1982,ferreira2006}), and because magneto-hydrodynamical (MHD)-type winds are an independent, extremely effective source of angular momentum transport \citep{blandford1982}, which may prove fundamental to reproduce accretion cycles in disk-driven systems \citep{scepi2018}.
Indeed, outflows are often realized in MHD simulations assuming certain initial configurations of the magnetic field (see e.g. \cite{zhu2018,jacquemin-ide2021} 
and references therein). 
Since it is difficult to discriminate between thermal and MHD winds of BHXBs from the line signatures, at least in CCD-level observations, the identification has often been made by elimination. Namely, the observed absorption lines are interpreted as evidence for magnetic winds when physical parameter values derived from the fitting to those lines are inconsistent with the predictions of the thermal winds. 
Such a magnetic wind candidate was first discovered in GRO J1655$-$40 \citep{miller2006,miller2008}, in which the distance of the wind $R$ was much smaller than that predicted for a thermal driven wind. \citep{fukumura2017} found that the observed feature can be well explained by the magnetic wind model assuming a self-similar magnetic field structure. However, subsequent studies have pointed out, based on detailed analyses of the spectral features and timing properties, that this wind can be explained as a Compton-thick, thermal (plus radiation-pressure) driven wind \citep{uttley2015,neilsen2016,shidatsu2016,tomaru2022}. 

A clear difference between thermal and magnetic winds is the wind velocity. As described above, thermal winds are only launched in the outer disk regions, while magnetic winds can be present at smaller radii where the higher escape velocities are reached thanks to magnetic torques.
%higher escape velocities are required. 
Therefore, resolving the absorption line profiles allows us to distinguish the wind launching mechanism, by revealing the presence or absence of faster velocity components and the profile of the line \citep{tomaru2022,chakravorty2023}. This was however impossible 
using X-ray CCDs, as they have a limited energy resolution around the Fe K band. Although high-velocity components were suggested using third-order Chandra grating spectra \citep{miller2015,miller2016,trueba2019}, spectral statistics were very limited. To settle this debate, observations with both higher photon statistics and superior energy resolution have long been awaited.

\section{XRISM/Resolve Spectroscopy of BHXBs}
\label{sec:xrism}

The launch of the X-ray satellite XRISM in 2023 has opened a new era in studies of winds in BHXBs. XRISM carries the new technology X-ray microcalorimeter named Resolve \cite{ishisaki2025}, which achieves an unprecedented energy resolution of $\sim$ 4.5 eV at 6 keV (on-orbit performance; \citep{kelley2025,porter2025}), together with 
a good photon-collecting power, both of which represent approximately an order-of-magnitude improvement over the Chandra grating in the Fe K band.  
This combination makes XRISM/Resolve the most powerful X-ray spectrometer currently available for investigating winds in BHXBs. 

Following the start of scientific operations in February 2024, XRISM has observed several BHXBs. In the following sections, we summarize the main results from these observations, and present them in order of spectral states, from the most to the least standard with respect to X-ray wind detections.
%XRISM also carries X-ray CCD named Xtend...

\subsection{4U 1630$-$472: Resolving Absorption Lines in the High/soft State}
4U 1630$-$472 has long known as a BHXB candidate with wind absorption lines (e.g., \citep{kubota2007,diaztrigo2014,hori2018,gatuzz2019, trueba2019,parra2025}). The system parameters, such as the orbital period, the black hole mass, the stellar type of the companion, are not determined 
(except for the distance to the source \cite{kalemci2025}), because of the very high Galactic absorption. The source exhibit outbursts 
generally repeating over $\sim 600$ days. XRISM observed the source on February 16--17, 2024, only $\sim$ 1 week after the start of its scientific observations. This was the first target-of-opportunity (ToO) observation of the XRISM mission. The source was in the last part of the decaying phase of a large outburst started from 2023. In the following, we summarize the results of the observation. We refer readers to \citep{miller2025} for details. 

\begin{figure}
    \centering
    \includegraphics[width=0.7\linewidth]{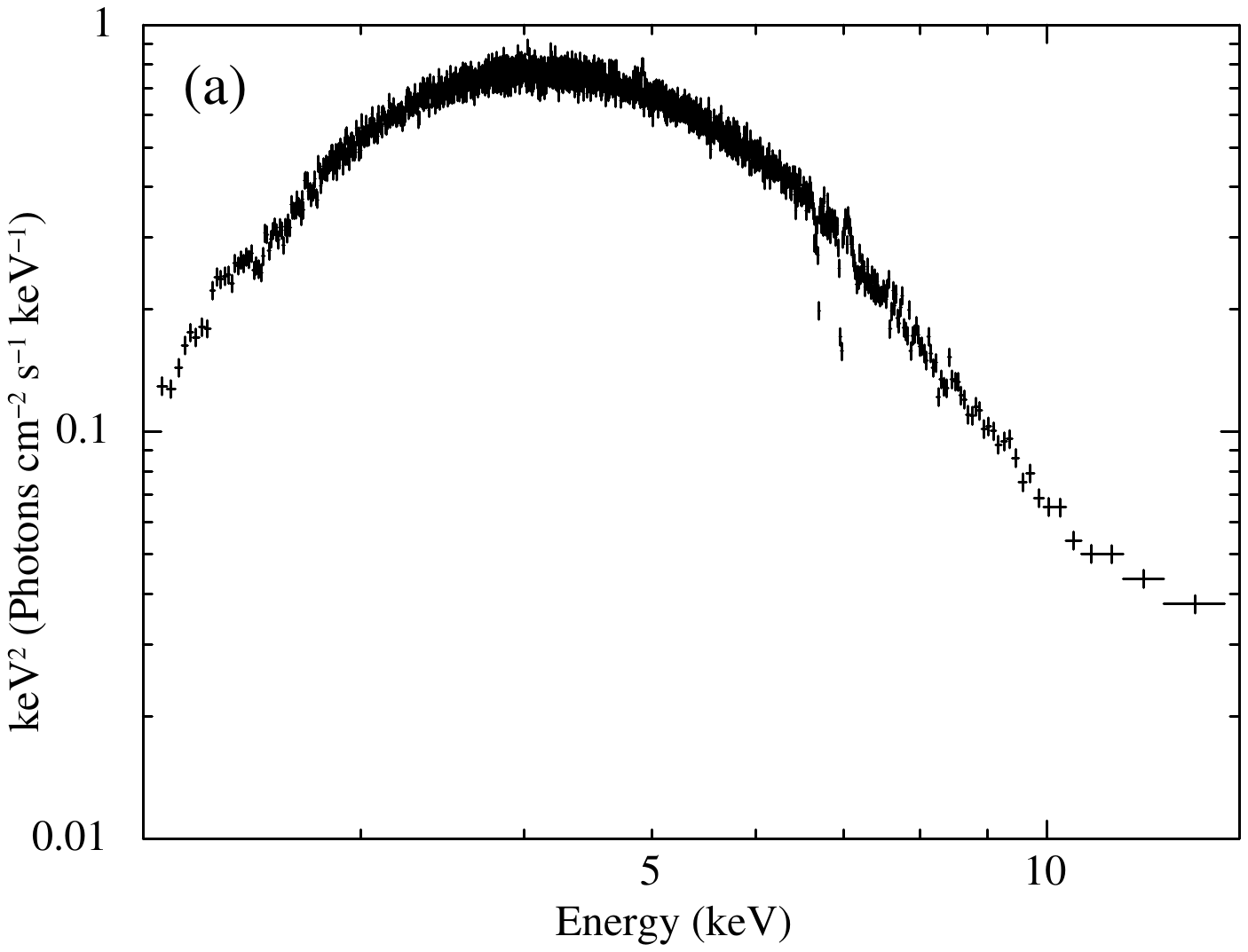}
    \includegraphics[width=0.7\linewidth]{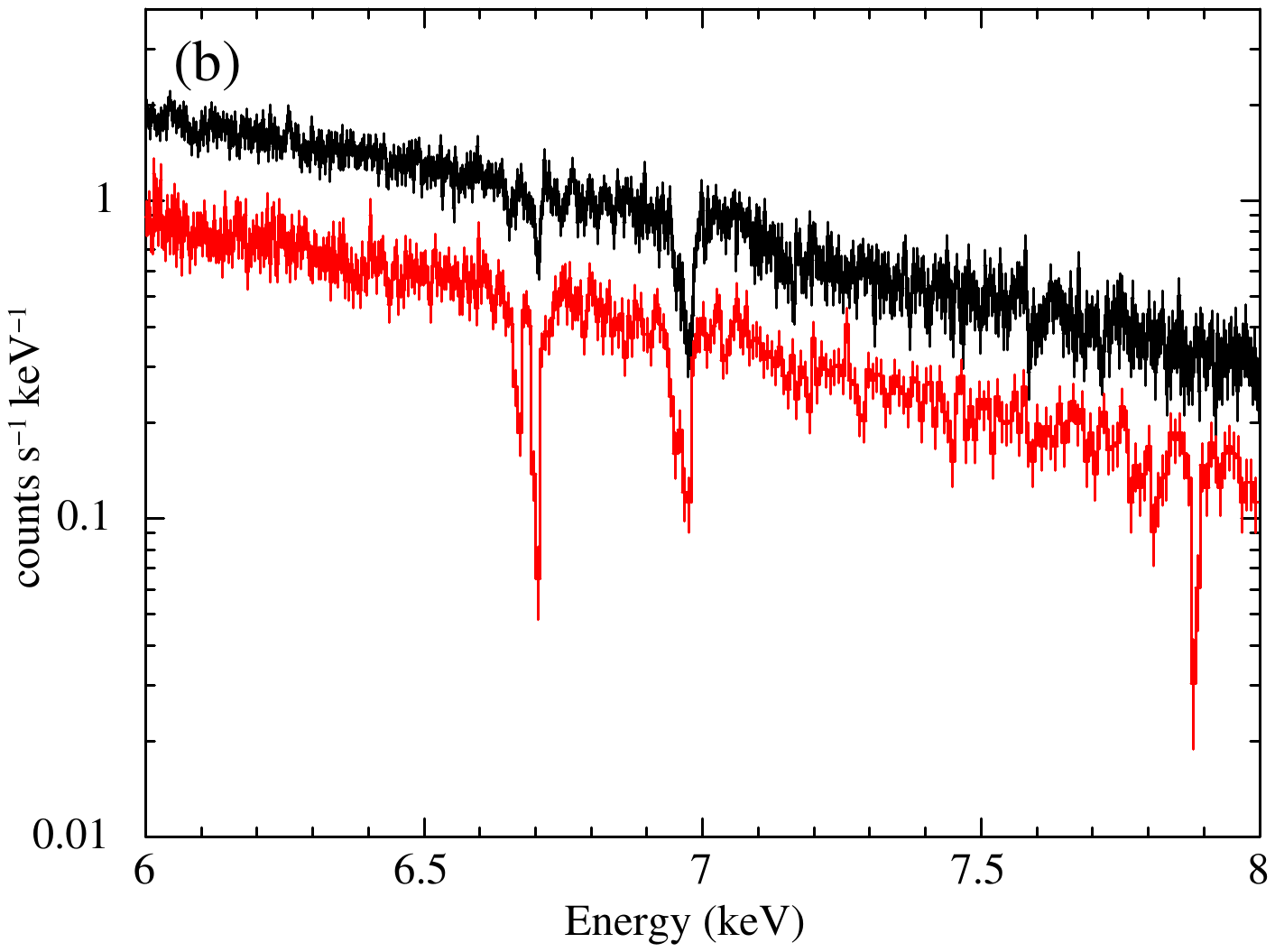}
    \caption{(a) Resolve time-averaged spectrum of 4U 1630$-$472. (b) Resolve spectra in the persistent phase (black) and the dip phase (red), around the Fe K band.}
    \label{fig:1630spec}
\end{figure}
Figure~\ref{fig:1630spec}(a) presents the time-averaged Resolve spectrum in 2--10 keV, which clearly shows a disk blackbody component, indicating that the source was still in the 
disk-dominant soft state. The unabsorbed bolometric luminosity was estimated to be $\sim 6 \times 10^{37}$ erg s$^{-1}$ or $\sim 0.05~(M_{\rm BH}/10~M_\odot)^{-1}~L_{\rm Edd}$ (where a  distance of 11 kpc is assumed \cite{kalemci2025}) using a multi-color disk and its thermal Comptonisation model. The source showed a flux drop by $\sim 10\%$ in the last part of the observation, which could be consistent a broad ``absorption dip'', seen in high inclination sources. Figure~\ref{fig:1630spec}(b) shows an enlarged view of the Resolve spectra at the persistent phase (outside the dip) and at the dip phase. In both phases, we successfully detected absorption lines, including the Fe XXV He$\alpha$ and Fe XXVI Ly$\alpha$ lines. This marks one of the lowest luminosity detections of ionized absorption lines ever achieved in BHXBs, and the profiles of these two lines were successfully resolved thanks to the high energy resolution and sensitivity of Resolve. These absorption lines were found to have complex profiles and require multiple zones of ionized absorber. The line-of-sight velocity of the main component dominating the two lines at the persistent phase was estimated to be only $\lesssim 200$ km s$^{-1}$, which is comparable to the orbital/systemic velocities and too small for the gas to escape from the gravity. This indicates that the lines are produced by a hot atmosphere bound on the accretion disk, rather than a wind. In the thermal wind scenario, the luminosity in this observation is likely lower than the critical luminosity to efficiently heat gas via Compton scattering and launch a wind, so thermal wind is not predicted, consistent with the observed absence of winds. 

Interestingly, the lines, especially Fe XXV He$\alpha$, became more stronger in the dip phase. This variation was found to be too large to explain by the change of the flux or spectral shape of the continuum. This indicates that the variation of lines is unlikely to be caused by the change in the illuminating X-rays but an intrinsic change in the properties of the absorber and are consistent with the decrease in flux of the continuum. This additional absorption component was found to have somewhat lower ionization and located at larger radii than the ionized absorption component persistently seen. We interpreted this additional component as a local structure, maybe created by the impact of the accretion stream from the companion star in the outermost disk region, passed our line of sight.

\subsection{MAXI J1744$-$294: a Detailed Look at a Galactic-center Source in the High/soft State}
The Galactic Center (GC) transient MAXI J1744$-$294 was discovered in January, 2025 by MAXI in the Galactic center region \citep{kudo2025}. Subsequent follow-up observations with X-ray observatories including MAXI, Swift, Chandra 
found that the source is likely a BHXB \citep{nakajima2025,heinke2025,mandel2025}. Although the radio counterpart was detected \citep{grollimund2025, michail2025}, optical and infrared observations have been hampered by the very high Galactic absorption and the most of the system parameters have not been determined. The disk inclination angle was constrained to be $38^\circ$--$71^\circ$ from the X-ray polarimetric observation with IXPE \citep{marra2025}. 

XRISM observation was conducted on March 3, 2025, as a Directors Discretionary Time (DDT) observation, when the source was in the high/soft state. 
The field of view of Resolve showed a blend of spectral features from the continuum of MAXI J1744$-$294, which dominated in the soft X-ray band, the neighboring neutron star AX J1745.6$-$2901, which dominated in the hard X-ray band, and the diverse sources of diffuse emission, most notably the supernova remnant Sgr A East and the Galactic center X-ray emission (GCXE), which dominated part of the main iron transitions. A comprehensive modeling of all sources in the FoV, incorporating the previous high-resolution view of the diffuse emission in the Galactic center \cite{xrism2025}, revealed a surprising diversity of emission features intrinsic to the BH spectrum. The most important and significant is the existence of a dense, highly ionized ($\log\xi \gtrsim 5.5$), static ($v_{\rm out}\lesssim -200$ km s$^{-1}$) emission layer, mainly expressed in the Fe XXVI and Fe XXV K$\alpha$ lines. 
This would be the first evidence for such emission features in non-obscured BHXBs with low mass companions, which historically have showed either absorption lines or a smeared reflection profile depending on their accretion state. It is thus possible that this source shows a "low-inclination" view of the standard wind/atmosphere features commonly seen in absorption in highly inclined sources. 

These lines would have remained undetected due to their weak equivalent width, and the limited spectral resolution and line detectability before the XRISM era.  
Future observations of low-inclination sources in similar accretion states will help confirm or rule out this possibility, but another likely scenario is that the source is a complete outlier in terms of line properties. This is reinforced by the detection of multiple additional narrow emission features between 6.70 and 7.1 keV, none of which match strong atomic transitions and can thus only be explained by strongly blueshifted or redshifted emission lines, with velocities up to $\sim-6000$ km/s (in the case of blueshifts) or down to $\sim 9000$ km/s (in the case of redshifts). 

This is even more unprecedented in binary systems, and the only case with somewhat similar features is the super-Eddington obscured source SS 433, with an 
optically-thin thermal X-ray spectrum showcasing oscillating features from relativistic, precessing jets (see e.g., \cite{shidatsu2025}). We thus attempted different physical modeling avenues, and compared photoionization and collisional ionization origins, as well as the physical interpretations of wind and jets origin. As the two mechanisms can achieve relatively similar line profiles, further analysis of the exhaustive broadband and multi-wavelength datasets available will help lift the degeneracy and rule out part of the proposed physical scenarios. More details of the XRISM background modeling, spectral analysis and physical modeling will be presented in forthcoming papers (Parra et al. in prep.). 

\subsection{IGR J17091$-$3624: High-resolution Spectroscopy in the Low/hard State}
%XRISM also observed the BHXB IGR J17091$-$3624. in its 2025 outburst. 
%This source 
IGR J17091$-$3624 is known to be a transient BHXB candidate with a high inclination angle, exhibiting outbursts every 2--3 years (e.g., \cite{kuulkers2003, capitanio2009, tetarenko2017}). Like other BHXBs, it usually shows state transitions between the low/hard state and the high/soft state in the outbursts \citep{capitanio2012}. Uniquely, it also occasionally enters rare variability states, including the so-called ``heartbeat'' oscillation state, similar to those seen in GRS 1915$+$105, in which ionized absorption lines are observed \citep{altamirano2011, wang2024}. 

We conducted a ToO observation of IGR J17091$-$3624 in February 2025, around the flux peak of the 2025 outburst. Nearly simultaneous observations were carried out with NuSTAR \citep{adagoke2025}, followed by an IXPE observation in March \citep{ewing2025}. The outburst eventually resulted in a so-called ``failed'' outburst, without a state transition, and the source remained in the low/hard state throughout the entire period. During the XRISM observation, the source showed a hard power-law shaped spectrum and strong short-term variability on timescales of seconds, confirming that it was in the low/hard state. We detected a broad Fe K emission line and a spectral hump around 20 keV, both likely produced by relativistic reflection in the inner region of the disk. No significant ionized absorption features were found in the Fe K band, suggesting that disk winds were either strongly suppressed or highly ionized at this epoch (Kobayashi et al. in prep). XRISM has also observed 1E 1740.7$-$2942 in the low/hard state, and no prominent absorption lines were detected in the Fe K band in this observation either.

\subsection{V4641 Sgr: Resolving Emission Lines from the Obscured Source} 
V4641 Sgr is also a Galactic transient BHXB exhibiting outbursts once per a few to several years (see e.g., \cite{tetarenko2017}). The source consists of a black hole with a mass of $\sim 6 M_\odot$ and a B9III companion with $\sim 3 M_\odot$, orbiting with a period of $2.82$ days \citep{macdonald2014}. It shows somewhat unusual flux and spectral evolution in its outbursts. The outburst in 1999 was a violent one, exhibiting a rapid flux change by a few orders of magnitude up to largely super-Eddington luminosities within a few hours \citep{smith1999,hjellming2000,inzand2001}. Subsequently, it showed much weaker outbursts with a typical peak luminosity of $\sim 10^{-3}~L_{\rm Edd}$, and the spectrum was often dominated by the unusual reflection-like component \citep{morningstar2014,koljonen2020}, or a disk blackbody-like component \citep{bahramian2015,pahari2015,shaw2022,connors2025}, which is usually observed at $\gtrsim 1$ order-of-magnitude higher Eddington ratios. In the 2020 outburst, narrow highly ionized emission lines, such as Fe XXVI Ly$\alpha$ and Fe XXV He$\alpha$, were detected with Chandra/HETG \citep{shaw2022}. These unusual spectral properties suggest that the inner disk regions are obscured by the outer disk or its dense atmosphere or wind, such that only a small fraction of the emission from the central source scattered into the line of sight, together with narrow emission lines from the irradiated outer regions, is observed.

A XRISM DDT observation of V4641 Sgr was conducted on September 30, 2024, when the source was in the last part of the decaying phase of its outburst. Despite the faintness and the short ($\sim 10$ ks) exposure, Resolve successfully detected narrow emission lines, including the highly ionized FeXXVI Ly$\alpha$ line and Fe XXV He$\alpha$ lines, as well as neutral Fe K$\alpha$ line. The continuum spectrum can be characterized with a multicolor disk blackbody model \citep{mitsuda1984}, with an inner disk temperature of $\sim 1.8$ keV. In this case, the luminosity was estimated to be $\sim 10^{35}$ erg s$^{-1}$ or $\sim 10^{-4}~L_{\rm Edd}$, which is orders of magnitude lower luminosity than the typical high/soft state, where a similar spectrum is observed. The low luminosity but the relatively high inner disk temperature resulted in an inner disk radius much smaller than that of the innermost stable circular orbit of the black hole.
The combination of the low luminosity and relatively high inner disk temperature implies an inner disk radius, much smaller than that of the innermost stable circular orbit of the black hole. This indicates that the central source was obscured as in previous observations and the intrinsic luminosity may be higher by orders of magnitude. The emission lines were found to have different Doppler shifts ranging $\sim \pm 1000$ km s$^{-1}$ and vary strongly on timescales of $\sim 10^3$, indicating a complex, inhomogeneous geometry of the obscuring high-density plasma.
We also conducted completely simultaneous optical spectroscopy using the 3.8m Seimei telescope, located at the Kyoto University Okayama Astronomical Observatory, Japan \citep{kurita2010}. We adopted the KOOLS-IFU spectrograph \cite{yoshida2005,matsubayashi2019} with the VPH-683 and VPH-495 grisms, which covers 5800--8000 \AA and 4300--5900 \AA with wavelength resolutions of $R/\Delta R\sim 2000$ and $\sim 1500$, respectively. Together with the absorption lines from the companion star, we detected a clear emission component in H$\alpha$ that could originate from the outer disk or a cold and dense wind/disk atmosphere, but no significant blueshifted ($\sim 1000$ km s$^{-1}$) strong absorption feature detected in previous outbursts \citep{munozdarias2018}. For more details, we refer readers to \cite{parra2025b}. 

Another obscured source, GRS 1915$+$105, was also observed with XRISM in 2024 October. Again, very complex emission line features were resolved \citep{miller2025b}. The Resolve spectrum showed He- and H-like of many elements as well as a radiative recombination continuum features. These results suggest that the central source was obscured by the irradiated, photoionized outer disk. 

\subsection{Summary of the Current Status}

\begin{figure}
    \centering
    \includegraphics[clip,trim=4cm 0.2cm 0.8cm 1cm,width=1.0\linewidth]{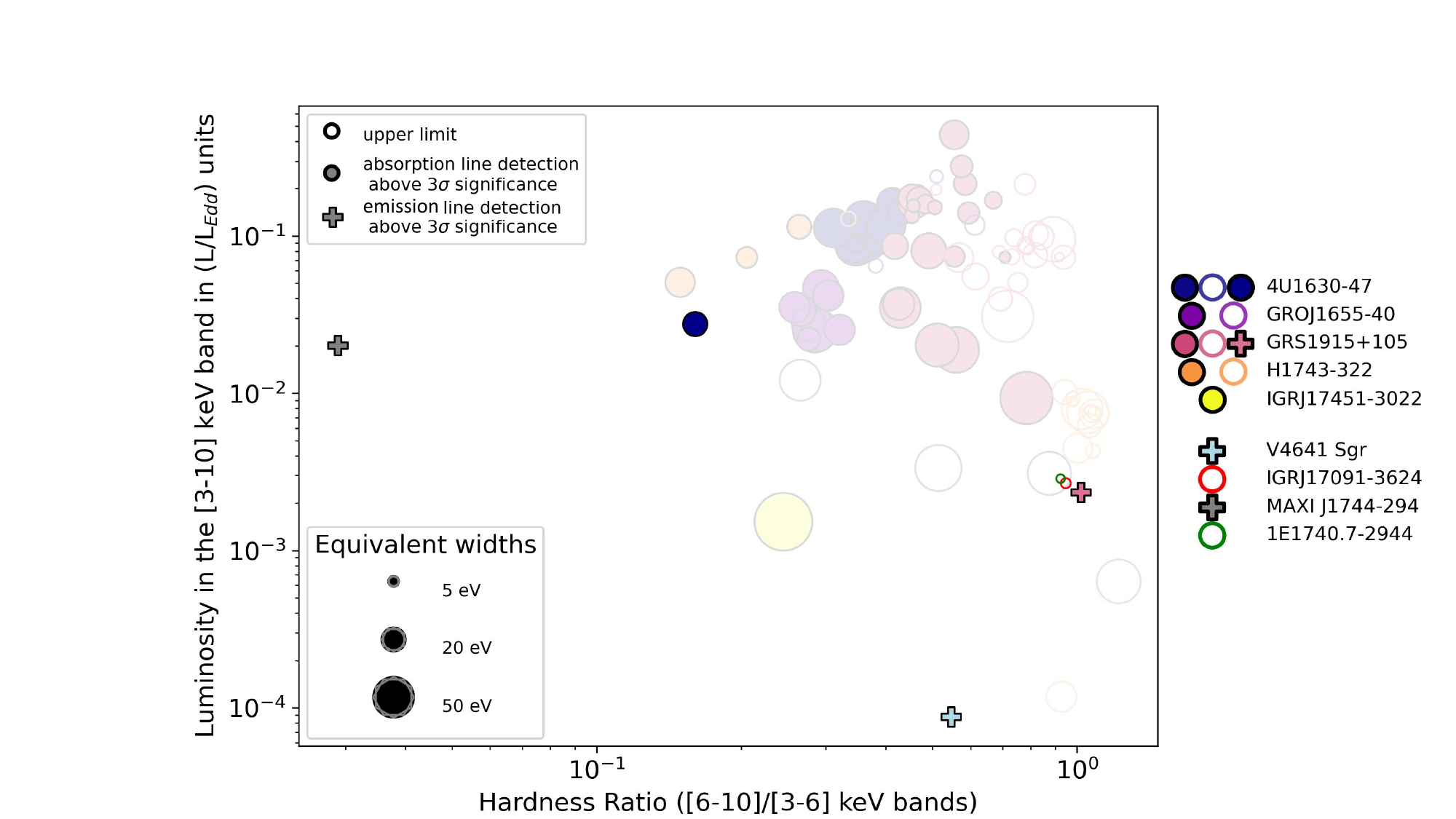}
    \caption{Hardness-luminosity diagram showcasing the position and line detections of all XRISM observations of BHXBs with low-mass companions performed as of fall 2025. Transparent markers show the subsample of BHXBs with wind detections in XMM-Newton and Chandra-HETG, from \cite{parra2024}. The source list is split between objects in that "identified" absorption line sample and the new XRISM observations. Emission lines markers are not scaled with respect to their equivalent width.}
    \label{fig:HLD_XRISM}
\end{figure}

%We have observed several BHXBs so far. 
We show in Fig.~\ref{fig:HLD_XRISM} the position of the different XRISM observations and their line detections within a hardness-luminosity diagram. 
In some observations, we successfully resolved spectral lines. They are, however, emission lines from unusual sources or absorption lines likely to be from a hot static atmosphere instead of winds. Although these observations are important to investigate the variety of wind properties and understand the evolution of winds with the luminosity and spectral states, more observations of BHXBs with high inclination angles are strongly needed, especially at bright phases in the high/soft state, where blueshifted absorption lines are usually seen. We illustrate this with the transparent markers in Fig.~\ref{fig:HLD_XRISM}, which show a sample of wind detections from the previous generation of instruments, from \cite{parra2024}, all of which are distant from current XRISM observations. 

\section{Summary and Future Prospects}
XRISM has begun to transform studies of disk winds in BHXBs by resolving absorption line profiles with unprecedented energy resolution and sensitivity in the Fe K band. 
At present, however, XRISM wind observations are still in their infancy, and only snapshots of several sources at limited epochs have been obtained. 
BHXBs with previous absorption line detections exhibit their outbursts only at most once per a few years. Moreover, the absorption lines are visible in a limited phase in outbursts and the visibility window of the XRISM satellite is also limited. Therefore, it will take long to reach a full understanding of the winds. 

As described in Section~\ref{sec:review}, absorption lines arising from winds evolve with spectral states, which reflect the inner disk structure. 
Jets are also thought to vary in response to the states, although their behavior is largely opposite to that of the absorption lines from winds. Understanding the physical origin of the disk–wind–jet connection is one of the key goals of wind studies in accreting black hole systems. To address this issue, spectral monitoring throughout outbursts is essential, particularly during state transitions from the low/hard state to the high/soft state. These transitions are of special interest because wind absorption lines begin to appear, while powerful transient jets are sometimes launched, providing a unique opportunity to study the interplay between accretion flows, winds, and jets.
In addition, spectral features from winds are not only seen in X-rays but also at lower energies, including the optical and infrared bands. This indicates the presence of cold winds, in addition to the hot winds seen in X-rays. In the last decade, studies of such cold winds has greatly progressed \citep{munozdarias2016,munozdarias2018,matasanchez2018,munozdarias2019, charles2019,panizoespinar2022,ambrifi2025}. They are seen in moderate inclination sources in which the spectral features of hot winds are usually absent and show a different state dependence from hot winds \citep{sanchezsierras2020}. 
The association between the hot and cold winds is an important yet unresolved issue. Addressing these issues requires observations beyond the X-ray band alone; coordinated multi-wavelength observations are therefore essential for a comprehensive understanding of winds in BHXBs. 

More recently, ultra-high-energy gamma-ray diffuse emission in the 1--100 TeV range has been detected in several BHXBs, suggesting that these systems may act as efficient particle accelerators  \citep{alfano2024,lasso2024}. 
In particular, the detection of gamma rays at energies $\gtrsim 100$ TeV implies that the parent particles are accelerated beyond 1 PeV, indicating that BHXBs could be potential PeVatrons, whose origin has long remained elusive. However, the mechanisms by which particles are accelerated in the vicinity of BHXBs are still poorly understood. Since some of the sources exhibiting ultra-high-energy gamma-ray emission are transient systems, it is crucial to clarify how strongly the central engine can influence its surroundings at different phases of an outburst. It is also important to search for diffuse signals in the X-ray band associated with gamma-ray signals. Along with Resolve, XRISM is also equipped with the X-ray CCD Xtend \citep{noda2025,uchida2025}, which has one of the largest field of view among current largest-class X-ray missions. The combination of its wide field of view and low instrumental background makes Xtend useful for this study \citep{suzuki2025}.

\end{document}